\documentclass[10pt, conference, compsocconf]{IEEEtran}
\usepackage[pdftex]{graphicx}
\graphicspath{ {./images/} }

\DeclareGraphicsExtensions{.pdf,.jpeg,.png,jpg}

\usepackage[cmex10]{amsmath}

\usepackage{array}

\ifCLASSINFOpdf
\else
\fi
\begin{document}
%
\title{e-Sem: Dynamic Seminar Management System for Primary, Secondary and Tertiary Education}

\author
{
	\IEEEauthorblockN{Ioannis A. Skordas \IEEEauthorrefmark{7},
		Nikolaos Tsirekas\IEEEauthorrefmark{1},
	Nestoras Kolovos\IEEEauthorrefmark{5},
		George F. Fragulis\IEEEauthorrefmark{1},
		  \\Athanasios G. Triantafyllou\IEEEauthorrefmark{7}   
		and
		Maria G. Bouliou\IEEEauthorrefmark{4}}\\

	\IEEEauthorblockA{\IEEEauthorrefmark{1}Laboratory of Web Technologies \& Applied Control Systems\\ Dept. Of Electrical Engineering\\
	Western Macedonia Univ. of Applied Sciences, Kozani, Hellas 
	\\ \IEEEauthorrefmark{7} Lab. of Atmospheric Pollution \& Environmental Physics, Dept. of Geotechnology and Environmental Engineering\\ Western Macedonia Univ. of Applied Sciences,\\ \IEEEauthorrefmark{5} Engineering Geologist \\
\IEEEauthorrefmark{4} Economist researcher }

}


\maketitle

\begin{abstract}
This paper describes the dynamic seminar management system named "e-Sem", developed according to the open-source software philosophy. Due to its dynamic management functionality, it can equally adapt to any education environment (Primary, Secondary, Tertiary). The purpose of the proposed dynamic system is ease of use and handling, by any class of users, without the need of special guidance. Also,  students are given the opportunity to: a) register as users; 
b) enroll in seminars in a simple way; c) receive e-learning material at any time of day any day of week, and d) be informed of new announcements concerning the seminar in which they are enrolled . In addition, the administrator and the  tutors have a number of tools such as : management seminars and trainees in a friendly way, sending educational material as well as new announcements to the trainees; the possibility of electronic recording of presence or absence of the trainees in a seminar, and  direct printing of a certificate of successful attendance of a seminar for each trainee. The application also  offers features such as electronic organization, storage and presentation of educational material, overcoming the limiting factors of space and time of classical teaching, thus creating a dynamic environment.

\end{abstract}


%
\IEEEpeerreviewmaketitle

\section{Introduction}
World wide web is undoubtedly a part of humans' everyday life. Conventional communication methods included reporting activity as well as radio, newspaper and television announcements. Nowadays the reading of news articles,  trip planning,  extraction of information from an encyclopedia and goods buying are indicative activities which many people deal mainly via the Web \cite{Berners} 

Any information can now be send over the Internet fast, with additional historical data through a website and in a simplified  way to the wide audience. Therefore, the combination of telecommunications and new technologies creates an evolving development framework system, making it more flexible with many abilities and more user-friendly \cite{Xuan}. 

There are many commercial Internet applications on the market which are subject to restrictions such as the high purchase value, maintenance cost, absolute dependence on the manufacturer, oligopolistic practices of some software companies (e.g. high annual utility costs, unexpected increases, withdrawal and lack of support) and the absence on the part of several manufacturers innovation  \cite{Sturgess}. 
Instead,  "Free Software / Open Source Software" (FOSS) provides significant advantages for users, giving them the freedom to run, copy, distribute, study, modify and improve the software without restrictions (Free Software Foundation, 1996- 2007) and have  numerous scientific applications (see \cite{SATEP}, \cite{Skordas_Fragulis_Triant2011}, \cite{Skordas_Fragulis_Triant2014}, \cite{Tele1}-\cite{Tele5} and the references therein.

Seminars Management Systems are sophisticated web applications, which are developed  by institutions or companies that want to work with digital learning (e-learning/distance learning), either to provide services to third parties or to train their staff \cite{Avgeriou}.

\section{System Description}
Several open seminar management software systems (SMS) are used today. The most popular(no. of users) are CoMPUS, e-Class and Moodle. Most SMS are composed of many individual parts, having in common most of the following \cite{Britain}:
\begin{itemize}
\item users in a seminar management system are divided into  students, tutors and administrators, and the access to the system is determined by the distinct role each one of them. 
\item SMS are Operating in a client-server model. 
\item they maintain some form of certification for users and divide them into groups, so that the same platform can be used for more than one seminar. 
\item they have a friendly interface for all users (students and  tutors). 
\item there is the ability to save information data about the user (creating profiles) and "help" during navigation of the application.
\item The environment works with web browsers so it is accessible from anywhere with any operating system, and is not necessary for users to install other software. \item make use of open source tools such as HTML, PHP, MySQL and Apache Server \cite{Britain}.

\end{itemize}
According to the above requirements,we propose "e-Sem" that is  a new dynamic seminar management system that can be utilized at primary, secondary and tertiary levels of education as well as at summer schools and other scientific/research communities. "e-Sem" is an application based on open source software tools, such as HTML, MySQL, PHP, JavaScript and JQUERY see \cite{Duckett}, \cite{web2008},\cite{Java1}, \cite{Binstock}, \cite{Atkinson},\cite{Vaswani}, \cite{mysql}, \cite{Chaffer}.

  The purpose of the proposed "eSem" management system is to provide trainees the ability to participate in online tutorials(seminars) in an easy, fast and comprehensible manner. In addition, all data are dynamically managed, while the necessary data validation checks are performed with automated actions.



The complete diagram of the system is shown in Figure 1, where three different user groups are supported. 
The first group are  "Trainees/students", who can  enroll in seminars, while they can see  a list of seminars taken in the past. In addition, they are given the opportunity to edit their profile and can receive educational materials and be informed about news and announcements for the specific seminar that are enrolled. The second group of users is the "Teachers/Tutors" with management rights (adding, deleting and modifying students) on seminars. It is also possible to record attendance, as well as to print a certificate on successful completion of a seminar for each trainee. The system automatically refuses to register a trainee in a seminar when the maximum number of participants are met. "Teachers/Tutors"  also have the ability to send educational material(lectures, exercises, exams), news and announcements to the trainees that participate in the specific seminar. Finally, the third group consists of the "Administrator" with the exclusive privilege of adding, deleting and modifying students, tutors and seminars. 

\begin{figure}[!h]
	\centering
	\includegraphics[scale=0.5]{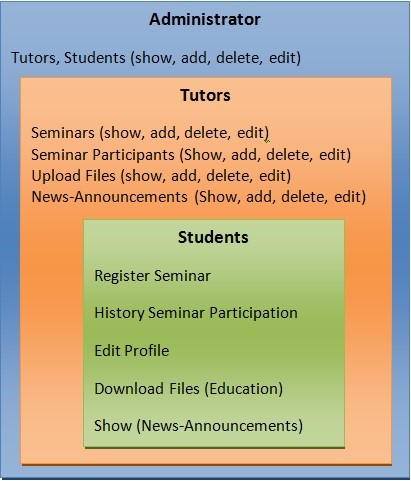}
	\caption{Diagram of «e-Sem» system}
\end{figure}

We describe  below  the functional and non-functional analysis requirements in details:\\
\textbf{Non-Functional:}

\begin{itemize}
	\item Operating in a client-server model
	\item There is no need for installing additional software as the environment works with web browsers so it is accessible from anywhere with any operating system.
\end{itemize}

\textbf{Functional:}	
\begin{itemize}
	\item The system supports three different user groups:
		\begin{itemize}
			\item The Trainees/students have the ability:
			\begin{itemize}
				\item enroll to seminars
				\item view their participation history
				\item edit their profile
				\item download educational material for seminars 
				\item show news - announcements
				\item print certificate of successfully completed seminars	
			\end{itemize}
			\item The Teachers/Tutors have the ability:
		\begin{itemize}
			\item manage seminars
			\item keep an attendance log for each of their seminars
			\item mark a student as a successful participant
			\item manage their seminars' participants 
			\item edit their profile
			\item upload educational material
			\item post news - announcements for their seminars	
		\end{itemize}
		
		\item The Administrator has the ability:
		\begin{itemize}
			\item manage seminars
			\item manage  seminars participants
			\item edit the profiles of  trainees and tutors
			\item add new tutors 
			\item post general  news-announcements
		 	
		\end{itemize}
		
		\end{itemize}	
\end{itemize}

\section {User Interface and Functionality}
This section describes the user interface, emphasizing the features and functions of the e-Sem application. As shown in Figure 2, there are two groups of users who have access to the following form (adding seminars). The first consists of tutors and the other one are the administrator. 
A number of validation checks are performed when entering data in order to add new seminars to the application.  In addition, it is possible to send educational material, as well as to delete, edit and display detailed information for each seminar.
\begin{figure}[!h]
\centering
\includegraphics[scale=0.3]{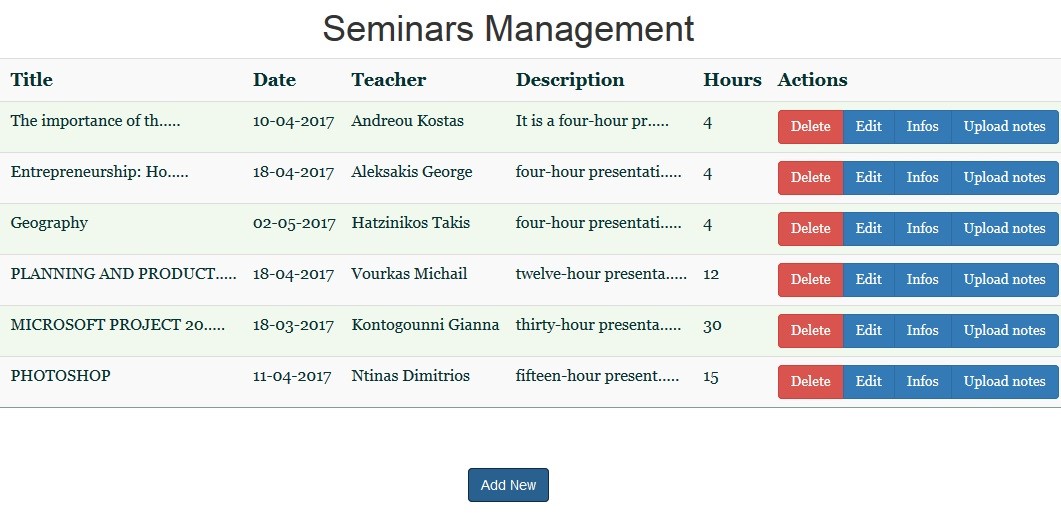}
\caption{Seminars' Management}
\end{figure}

Another basic feature of the e-Sem application  is that the teacher/administrator can check the attendance/absence of participants at a seminar in a hourly basis. The advantage in this case is that the successful / unsuccessful completion of the seminar for each trainee is easily and automatically determined (Figure 3).

\begin{figure}[!h]
\centering
\includegraphics[scale=0.3]{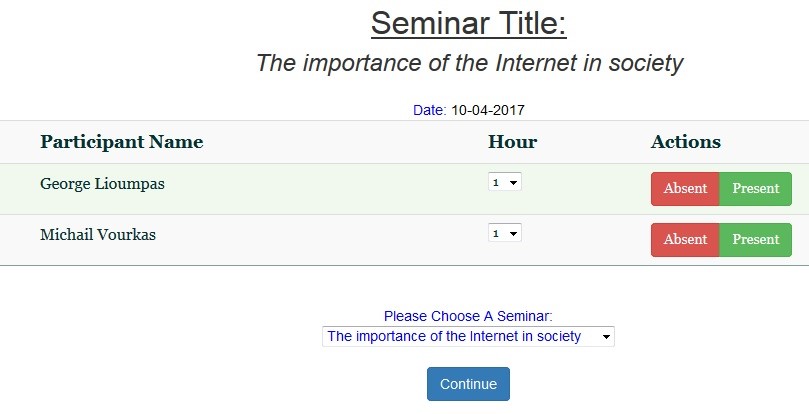}
\caption{Management of students attendance}
\end{figure}

The trainee has the ability to enroll in a seminar through a dynamic form that inform the trainee about the available seminars, while at the same time he/she is informed about the number of participants who have expressed interest so far (Figure 4) .

\begin{figure}[!h]
\centering
\includegraphics[scale=0.3]{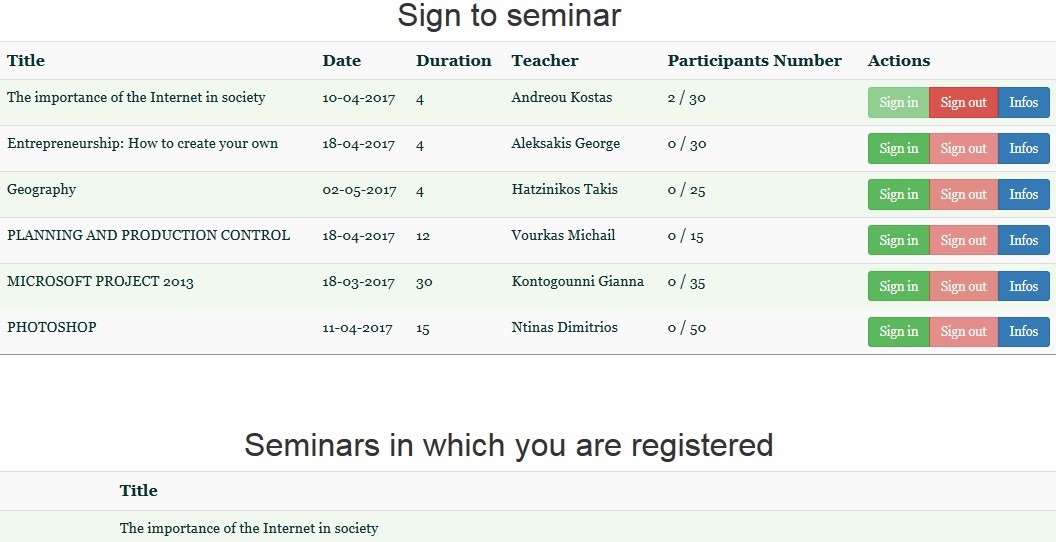}
\caption{Registration of seminars by trainees}
\end{figure}

As shown in Figure 5, past history of seminars is an additional feature that informs the trainee in which seminars has successfully participated in the past. Also  has the ability to directly obtain a certificate of successful attendance at any completed seminar.

\begin{figure}[!h]
\centering
\includegraphics[scale=0.3]{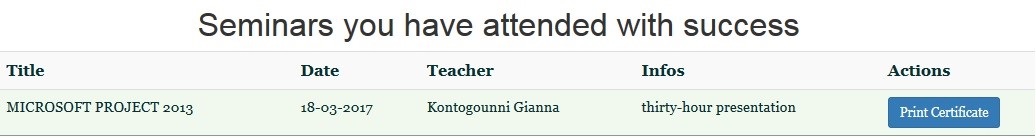}
\caption{Successful student attendance at seminars}
\end{figure}

 The teacher/administrator can send educational material in the form of a file to the participants of each seminar (Figure 6).
 
\begin{figure}[!h]
	\centering
	\includegraphics[scale=0.3]{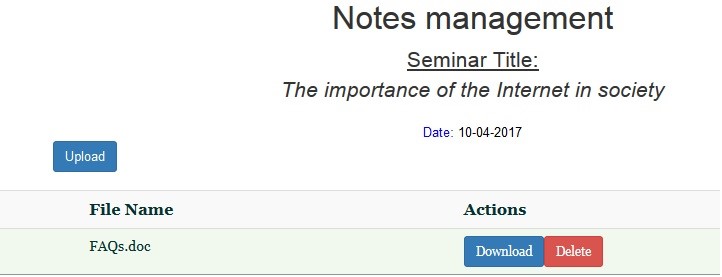}
	\caption{Sending educational material to trainees}
\end{figure}

Another function as shown in Figure 7 is to send new messages and announcements to specific tutors or  trainees/students, or multicast to everyone. This action can be done by the  tutors as well as by the administrator.

\begin{figure}[!h]
	\centering
	\includegraphics[scale=0.3]{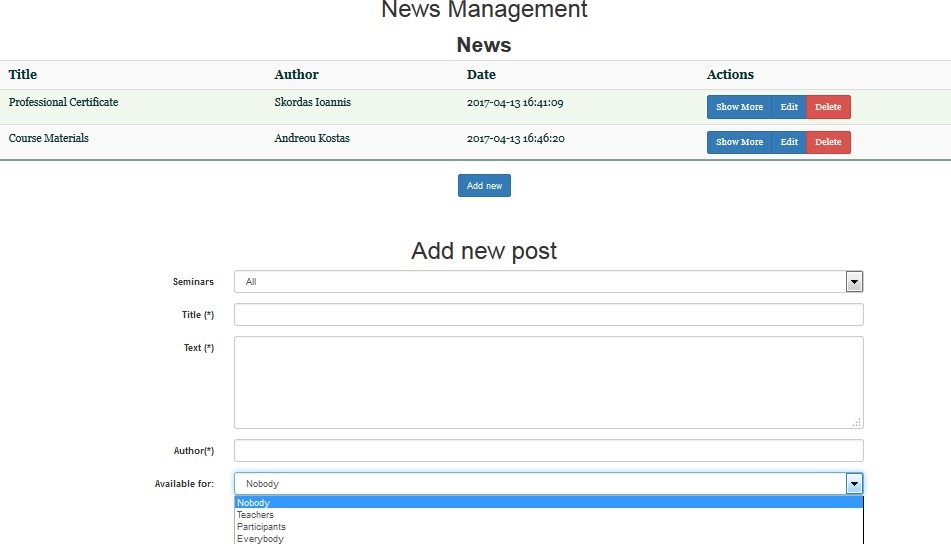}
	\caption{Sending new announcements to  tutors/trainees}
\end{figure}


In order to evaluate the functionality and effectiveness of the application, the following were performed:
The application was uploaded in a web server and test data was imported. 
Then we test the application's speed/response . The results were very satisfactory as the application worked fast enough. Then, we created an account for each user  and ask of a number of users to test it . These users come from various social groups and professions (developers, analysts, teachers, students, pupils). Then we briefly informed these users(by e-mail) about the subject that our application deals with. We have not provided detailed usage instructions to see if our application is functional and easy to use.
The comments we received were quite satisfactory and encouraging for the work we had done. Most test-users thought the application was very easy to handle and its operations quite clear. Most of the negative comments were about the visual approach of the application. The majority of these were taken into account and we took the necessary steps to improve them. In general, we would say that the feedback of our test users  helped us to improve several parts of the application.

\section{MySQL and Database Architecture}

MySQL is the most well-known Database Management System \cite{mysql}. It is an open source tool and its  advantage is that it is constantly improved in regular time intervals. As a result MySQL has evolved into a fast and extremely powerful database. It performs all functions like storing, sorting, searching and recall of data in an efficient way. Moreover, it uses SQL, the standard worldwide query language. Finally, it is distributed free. The Database which is used in the application
has been designed to use the storage engine InnoDB, so that the restrictions of the foreign keys can be created.  
For the needs of the web application «e-Sem», a database is build, that stores the necessary information in order to refresh dynamic content of the website, each time the data base are modified. In this way, it becomes easier to manage and view the contents of the application. The database called dbseminars with totally eight tables is shown in Figure 8.

\begin{figure}[!h]
	\centering
	\includegraphics[scale=0.45]{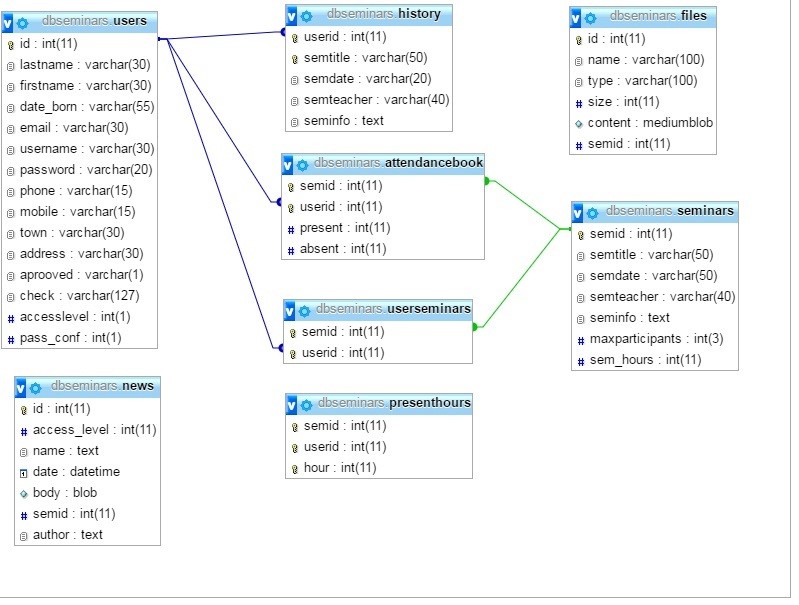}
	\caption{Architecture of Database «e-Sem»}
\end{figure}

The two main tables of the application are the users and the seminars, where the first consists of 15 fields with id key. One of the most important fields is "accesslevel”, which accepts only three values (0, 1 and 2). Depending on the value the user identification takes place, ie zero (0) for the admin, one (1) for the tutor and two (2) for the trainees/students. The remaining fields are the user's details when they are first logged into the system. The seminars table has as its main purpose to store seminars, having as a master key the “semid” that uniquely characterizes each entry. In addition, the “maxparticipants” field represents the maximum number of participants for each seminar. In the users seminars table, the students participating in the seminars are stored with the “userid” and “semid” as foreign keys and the key combination of them. The “attendancebook” has the role of recording attendance-absence during the attendance of seminars for each student. The main key is the combination of “semid” and “userid”, with foreign keys the corresponding fields. The “history” table represents the history of a successful student's participation in a seminar. As a foreign key is the “userid” and primary key the combination of “userid” and “semtitle”. The “news” table stores news and announcements, and consists of seven fields with id key. In the “files” table information such as the name, type, and the size of a file associated with the training material are stored. The key is id. Finally, the “presenthours” table saves the information of what time it is stated as present or absent in the particular student's seminar. The main key is the combination of the three “semid”, “userid” and “hour” fields.

\section{Conclusion}
In the present paper we propose a Seminars' Management System (e-Sem) with a dynamic environment that meets the needs of several levels of education (Primary,Secondary,Tertiary,summer schools etc.), developed with open source software tools and taking into account their advantages and features. Students/trainees can easily access  the system in a friendly and simplified way, while  tutors have  a management tool with many capabilities and functions.
Our application  use recent client side technologies as CSS3, bootstrap v3.3.7 framework for texts, forms, buttons, tables and navigation and  jQuery for JavaScript plugins.

 Extending  the abilities of the  e-Sem system is under developement from our research team .Specifically we work to add the following items (among other stuff):
\begin{itemize}
\item ability to representation the educational material in multimedia form such as  video and audio 
\item provide a live streaming ability
\item add a live chat room, a shoutbox  and a discussion forum for all users. 
\end{itemize}
Also another possible improvement of the proposed application could be the use of a Model View Controller (MVC) framework in order its modules to be better organized.

Finally we give the differences as well as the similarities between our application and existing ones as Moodle, Compus, e-class:\\
\textbf{Similarities}
\begin{itemize}
	\item Client-server technology
	\item Search for educational material
	\item Frequently Asked Questions (FAQ)
	\item News - Announcements
	\item File exchange	
\end{itemize}

\textbf{Differences}
\begin{itemize}
	\item Automatically send emails to registered users, when introducing announcements and educational material by administrator.
	\item Insert future events from all application users
	\item Multimedia
	\item forum and  live chat 
\end{itemize}


\section*{Acknowledgment}

Dr. George F. Fragulis work is supported by the program Rescom 80159 of the Western Macedonia Univ. of Applied Sciences, Kozani, Hellas.


%

\end{document}